\pdfoutput=1
\documentclass[%
 reprint,
 amsmath,amssymb,
 nofootinbib,
 aps,
 floatfix,
]{revtex4-1}

\usepackage[utf8]{inputenc} 

\usepackage{graphicx}
\usepackage{dcolumn}
\usepackage{bm}
\usepackage{hyperref}

\newcommand{\ex}{\mathrm{e}}
\newcommand{\im}{\mathrm{i}}
\newcommand{\Tr}{\mathrm{Tr}}

\usepackage{color}

\begin{document}

\preprint{APS/123-QED}

\title{Center clusters in full QCD at finite temperature and background magnetic field}

\author{G.~Endr\H{o}di}
\affiliation{Institute for Theoretical Physics, Universit\"at Regensburg, D-93040 Regensburg, Germany}
\author{A.~Sch\"afer}
\affiliation{Institute for Theoretical Physics, Universit\"at Regensburg, D-93040 Regensburg, Germany}
\author{J.~Wellnhofer}
\email{jacob.wellnhofer@ur.de}
\affiliation{Institute for Theoretical Physics, Universit\"at Regensburg, D-93040 Regensburg, Germany}


\begin{abstract}
We study the center structure of full dynamical QCD at finite temperatures 
and nonzero values of the background magnetic field using 
continuum extrapolated lattice data. 
We concentrate on two particular observables characterizing 
center clusters: their fractality and the probability for percolation. 
For temperatures below and around the transition region, 
the fractal dimension is found to be significantly smaller 
than three, leading to a vanishing mean free path inside the 
cluster structure.
This finding might be relevant for center symmetry-based models 
of heavy-ion collisions. 
In addition, the percolation probability is employed to define the 
transition temperature 
and to map out the QCD phase diagram in the magnetic field-temperature 
plane.
\end{abstract}

\maketitle

\section{Introduction}
\label{sec:Introduction}

Quantum chromodynamics (QCD) is the theory describing strongly interacting 
matter. 
QCD predicts the existence of a finite temperature transition that 
separates the low-energy confined regime and the deconfined quark-gluon-plasma (QGP)
phase. The properties of this transition are relevant for the evolution of the early universe
and are also probed by contemporary heavy-ion collision experiments, both at RHIC and at the LHC. 

Following the conjecture that 
the deconfinement transition in the gluonic sector is related to 
the magnetic transition of a corresponding spin system~\cite{Svetitsky:1982gs,Yaffe:1982qf}, 
and the finding that the latter 
can be understood in terms of cluster percolation~\cite{Fortuin:1971dw},
it was proposed that the gluonic field configurations of QCD can be characterized 
by center clusters and that the deconfinement transition may be understood as a percolation 
phenomenon~\cite{Bhattacharya:1990hk}. 
In this description 
confinement manifests itself in small and uncorrelated clusters, 
while the deconfined 
regime  exhibits a large cluster that percolates and induces long-range correlations.
The center structure of the QGP was also incorporated in models of heavy-ion collisions~\cite{Gupta:2010pp,Asakawa:2012yv}
and was argued to explain various properties of the plasma phase including 
its low shear viscosity and high (color) opacity~\cite{Asakawa:2012yv}. 
The main ingredient in this kind of models is the scattering 
of partons on the cluster walls, characterized 
by a mean free path. 

Besides the temperature, another parameter relevant for heavy-ion phenomenology is the 
background (electro)magnetic field generated by spectator particles in off-central 
collisions. Strong magnetic fields are also thought to have existed in the early stages 
of the universe and thus their effects on the QGP 
are of interest for cosmology as well. For recent reviews on 
the role of magnetic fields for strongly interacting matter 
see, e.g., Refs.~\cite{Kharzeev:2013jha,Andersen:2014xxa}.

In this paper we perform numerical lattice simulations to 
study center clusters in $2+1$-flavor QCD and determine their response 
to nonzero temperatures and background magnetic fields. 
We confirm that the clusters are not three-dimensional objects 
but instead have a fractal nature, 
as has already been observed in pure gauge theory (see, e.g., Ref.~\cite{Endrodi:2014yaa}). 
We demonstrate that as a consequence of this fractality the mean free path 
inside the clusters vanishes for temperatures and magnetic fields relevant for 
heavy-ion phenomenology. 
Furthermore, we propose a new observable for determining the transition temperature 
in full QCD and use it to map out the phase diagram in the magnetic field-temperature 
plane.

\section{Center clusters}
\label{sec:Preliminaries}

The concept of center clusters relies on the center symmetry of \emph{pure gauge} 
theory, formulated in Euclidean space-time at a nonzero temperature $T$.
Center symmetry denotes the invariance of the action under 
topologically non-trivial transformations $g$. These -- 
unlike normal gauge transformations -- are only periodic up 
to a constant twist, $g(x,t+1/T)=z\,g(x,t)$ 
in the Euclidean time-like direction~\cite{'tHooft:1977hy}. 
Here, $z$ belongs to the center
\begin{align}
\mathbb{Z}_3 = \{1,\ex^{-2\pi \im/3}, \ex^{2\pi \im/3}\},
\label{eq:center}
\end{align}
of the gauge group $\mathrm{SU}(3)$. 
While the confined phase is center symmetric for pure gauge theory, 
in the deconfined phase
this symmetry is spontaneously broken. The corresponding 
order parameter is the expectation value of the Polyakov loop, 
defined on the lattice as
\begin{align}
    P
    &=
    \frac{1}{V}\sum_x \Tr
    \prod_t U_4(x,t),
    \label{eq:avg_ploop}
\end{align}
where the non-Abelian vector potential $A_\mu$
is represented by group elements $U_\mu=e^{iaA_\mu}$,
and $V$ denotes the spatial volume of the system. 
For pure gauge theory the expectation value of $P$ 
vanishes below the transition temperature $T_c$ 
and selects one of the center sectors~(\ref{eq:center}) above the transition. 
In pure $\mathrm{SU}(3)$ gauge theory this deconfinement transition 
is of first order~\cite{Kogut:1982rt,Celik:1983wz}.

The presence of dynamical quarks modifies this picture slightly: 
the fermion determinant breaks center 
symmetry explicitly and always favors the 
trivial center element $1$ (see, e.g., Ref.~\cite{Kovacs:2008sc}).
However, 
this explicit breaking is rather mild 
and the Polyakov loop can still be used as 
an approximate order parameter. The corresponding deconfinement 
transition is no real phase transition but merely 
an analytic crossover~\cite{Aoki:2006we,Bhattacharya:2014ara}.
For a pedagogical introduction to center symmetry and the 
Polyakov loop, see Ref.~\cite{Holland:2000uj}. 

Although the expectation value of $P$ is -- due to the explicit 
breaking -- always real, it turns out that 
there are local domains in space, in which the Polyakov loop points towards 
one 
of the three center sectors~\cite{Fortunato:1999wr,Fortunato:2000fa,
Gattringer:2010ms,Gattringer:2010ug,
Dirnberger:2012gn,
Borsanyi:2010cw,Danzer:2010ge, 
Schadler:2013qba,Stokes:2013oaa}.
The corresponding {\it local} Polyakov loops $L(x)$ read 
\begin{align}
    L(x)
    &=
    \Tr
    \prod_t U_4(x,t),  \quad\quad
    P=\frac{1}{V} \sum_x L(x).
    \label{eq:local_ploop}
\end{align}

\begin{figure}[b]
        \centering
        \includegraphics[width=0.4\textwidth]{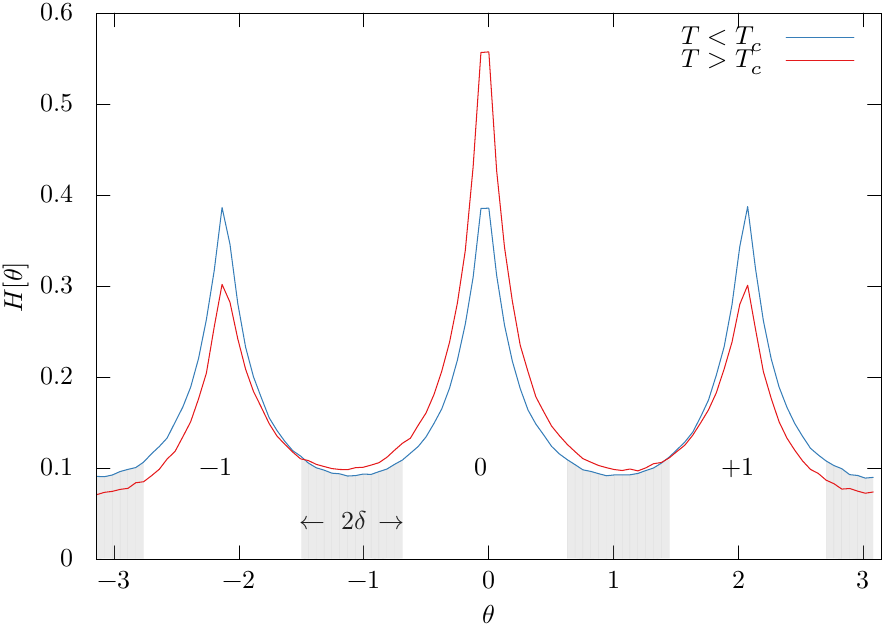}
        \caption[
            Illustration of center cluster definition.
        ]{Histogram of the local Polyakov loop phase below and 
            above the transition temperature and the definition 
            of sector numbers according to Eq.~\protect\ref{eq:sector_definition}
        }
        \label{fig:center_sector_def}
\end{figure}

Below $T_c$, all three sectors are (almost) 
equally represented, giving rise to a cancellation and an (almost) 
vanishing average Polyakov loop $P$. This is visualized in Fig.~\ref{fig:center_sector_def}, where the histogram of the 
local phase $\theta(x) = \arg L(x)$ is shown for a typical 
low-temperature configuration. For temperatures above $T_c$, the real 
sector $\theta\approx 0$ becomes dominant 
(also included in Fig.~\ref{fig:center_sector_def}) and induces a large real 
average Polyakov loop. 
This picture of center clusters has been studied in pure gauge 
theory with two~\cite{Fortunato:1999wr,Fortunato:2000fa},
with three~\cite{Gattringer:2010ms,Gattringer:2010ug,Endrodi:2014yaa} 
and with four colors~\cite{Dirnberger:2012gn}, while preliminary results 
for dynamical quarks have been obtained 
in Refs.~\cite{Borsanyi:2010cw,Danzer:2010ge}. 
(For visualisations of the clusters, see
Refs.~\cite{Schadler:2013qba,Stokes:2013oaa}.)
We mention that while the change in the distribution of $\arg L(x)$ is essential 
for the deconfinement transition, the modulus $|L(x)|$ was found to play no relevant role in this 
respect~\cite{Fortunato:1999wr,Fortunato:2000fa,
Gattringer:2010ms,Gattringer:2010ug,
Dirnberger:2012gn,
Borsanyi:2010cw,Danzer:2010ge, 
Schadler:2013qba}.

Besides the distinct population of the three sectors below and above $T_c$, 
there is another pronounced difference between the confined and deconfined 
regimes. While the clusters are small below $T_c$, they percolate and span 
across the total volume above the transition region. In this sense the deconfinement 
transition becomes very similar to the percolation phenomenon in a 
three-state spin system.
To give the center clusters a precise definition
that conforms to this picture, we need to impose a filter on the local phases 
$\theta(x)$ that discards sites lying far from center elements.
Specifically, to each site $x$ we
assign a sector number $n(x) \in \{-1,0,1\}$
in the following manner~\cite{Gattringer:2010ms}:
\begin{align}
    n(x)
    =\left\{
        \begin{array}{rlrcl}
            +1 & 
            \mathrm{ for } &
            \theta\in
            [\phantom{-}\frac{\pi}{3} +\delta\ ,
            \phantom{-}\pi-\delta], 
            \\
            0 & 
            \mathrm{ for } &
            \theta\in
            [-\frac{\pi}{3} +\delta\ ,
            \phantom{-}\frac{\pi}{3} -\delta ],
            \\
            -1 & 
            \mathrm{ for } &
            \theta\in
            [-\pi +\delta\ , 
            -\frac{\pi}{3}-\delta ],
        \end{array}
    \right.
    \quad \delta=\frac{\pi}{3}\cdot f.
    \label{eq:sector_definition}
\end{align}
Here, $f\in [0,1)$ is a free parameter, 
which removes ``undecided'' sites, i.e.,\ those that lie close to the minima 
of the distribution $H(\theta)$, see Fig.~\ref{fig:center_sector_def}.
In the following we will refer to $f$ as the 
cut parameter.
The center clusters are then constructed in the following way:
two neighboring sites $x$ and $y$ belong to the same cluster 
if their sector numbers are the same, 
that is, if $n(x)=n(y)$. 
This divides space into domains where the local Polyakov loop 
points towards one of the three center elements. 

We emphasize that a nonzero cut parameter is necessary 
to interpret the deconfinement transition as a percolation phenomenon.
Indeed, at $f=0$, the center 
clusters would percolate already at low 
temperatures\footnote{To see this, note that in random percolation theory, the critical 
probability for a three-dimensional cubic lattice is 
$p_c\approx 0.31<1/3$~\cite{Aharony2003}. Thus, even if the local Polyakov loops
are completely random (i.e.,\ the center sectors are equally populated) 
such that $p=1/3$, each of the three sectors will percolate on average.
For an explicit demonstration of this effect
see Refs.~\cite{Gattringer:2010ms,Endrodi:2014yaa}.
}.
By introducing $f\neq0$ and discarding sites lying far from center elements, 
the clusters are made thinner and percolation is delayed to set in only around $T_c$. 
This way, 
the confined phase exhibits clusters with finite size,
while in the deconfined phase there is one percolating cluster, 
as was demonstrated in pure gauge 
theory~\cite{Gattringer:2010ms,Gattringer:2010ug,Endrodi:2014yaa}.
Note that similar thinning techniques (cf.\ Ref.~\cite{Fortuin:1971dw}) 
to reduce the cluster size are necessary 
in different contexts as well, e.g., for the magnetic 
transition in the Potts model~\cite{Coniglio:1980zz} or for the 
droplet description of the Ising model~\cite{Fortunato:2000fa}.

\section{Results}
\label{sec:Results}

The results presented below are based on the gauge configurations 
generated in Refs.~\cite{Bali:2011qj,Bali:2012zg,Bali:2014kia,Endrodi:2015oba} at various 
values of the temperature, of the magnetic field $B$
and of the lattice spacing $a$. 
These ensembles have been produced using the Symanzik tree-level improved 
gauge action and $2+1$ flavors of stout smeared rooted staggered quarks 
with physical masses. 
Details of the 
simulation setup and of the algorithm can be found in 
Refs.~\cite{Aoki:2005vt,Borsanyi:2010cj,Bali:2011qj}.
In the following we consider the stout smeared gauge links for calculating the local Polyakov loops.

The vacuum configurations (corresponding to $T\approx0$, $B=0$) with several lattice 
spacings are used to set the cut parameter $f(a)$ in a consistent manner. 
At finite temperatures we consider $N_s^3\times N_t$ lattices and employ 
the fixed-$N_t$ approach to vary the temperature. That is to say 
the temperature $T=(N_ta)^{-1}$ is changed by tuning the lattice spacing $a$ for 
a fixed lattice geometry. 
In this approach the continuum limit corresponds to the limit 
$N_t\to\infty$ at a given temperature.
The magnetic field is chosen to point in the $z$-direction and 
enters the simulation setup via its quantized 
flux,
\begin{equation}
    \Phi=eB \cdot (aN_s)^2 = 6\pi N_b, \quad\quad N_b\in\mathbb{Z},
 \end{equation}
where the magnetic field is measured in units of the 
elementary charge $e>0$.
Due to flux quantization, an 
interpolation of the data at fixed $N_b$ 
is necessary to obtain results as a function of $eB$. 
For further details on the implementation of the magnetic field 
see Ref.~\cite{Bali:2011qj}.

\subsection{Scale setting}
\label{sec:Tuning_the_cut_parameter}

To set the cut parameter unambiguously additional physical input is necessary.
A possible way to set $f$ is to prescribe the value that
the physical radius $R$ of the largest cluster should take at low temperatures~\cite{Gattringer:2010ug,Endrodi:2014yaa}. 
For a cluster of size $s$, we define the radius $R$
by the mean squared deviation of the 
sites $\bold{r}_i$ in the cluster from its center of mass $\bold{R}_{\rm CM}$: 
\begin{equation}
    \bold{R}^2 = \frac{1}{s} \sum_{i=1}^{s} (\bold{r}_i - \bold{R}_{\rm CM})^2. 
 \end{equation}

\begin{figure}[t]
    \begin{minipage}[c]{0.4\textwidth}
        \centering
        \includegraphics[width=\textwidth]{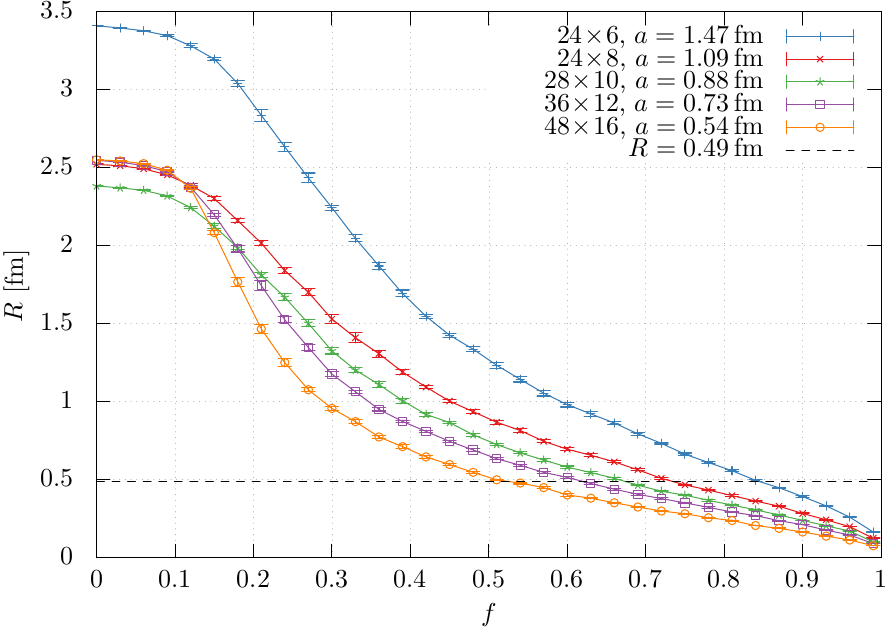}
        \caption[
            Radius of largest cluster vs cut parameter, QCD, smeared.
        ]{The radius $R$ of the largest cluster as a function of the cut parameter $f$ for our zero-temperature ensembles with 
                various lattice spacings. The dashed line indicates 
                the prescribed cluster radius $R=0.49\,$fm.
        }
        \label{fig:f_vs_r_phys_S}
    \end{minipage}
\end{figure}

To put this implicit prescription into practice we need to search 
for the value of $f$ where the largest cluster has the desired radius.
This procedure is visualized in Fig.~\ref{fig:f_vs_r_phys_S} for various 
zero-temperature lattice ensembles with different lattice spacings $a$.
Reading off the intersection of the $R(f)$ curves with 
the prescribed radius of $R=2.5\,\frac{1}{\textmd{GeV}}=0.49\, \textmd{fm}$ determines the scaling
relation $f(a)$.
Note that for $f\to1$, all sites are removed and, thus, 
the radius shrinks to zero, while for $f=0$, the largest cluster fills 
the total volume so that $R$ equals half the linear lattice size.
The value $R=0.49\textmd{ fm}$ which we chose for setting $f$ 
corresponds to a typical hadronic size
relevant for the low-temperature confined regime.
Note, however, that we are free to choose different radii as well. 
The subsequent analysis is performed using various 
values $0.35\textmd{ fm}<R<0.5\textmd{ fm}$.

\begin{figure}[b]
    \begin{minipage}[c]{0.4\textwidth}
        \centering
        \includegraphics[width=\textwidth]{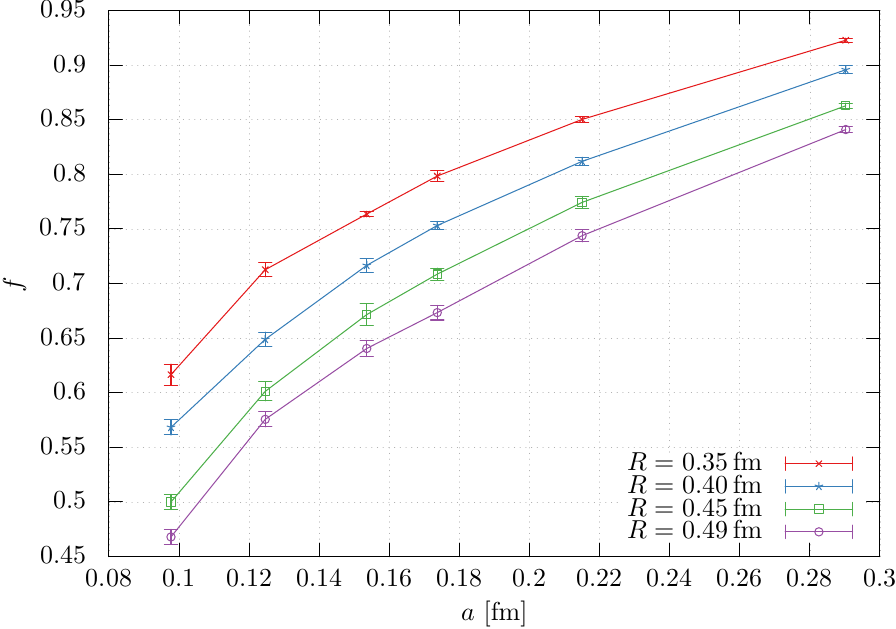}
        \caption[
            Cut parameter vs $\beta$, QCD.
        ]{The cut parameter as a function of the lattice spacing
                for various fixed cluster radii.
        }
        \label{fig:beta_vs_f}
    \end{minipage}
\end{figure}

The so obtained dependence $f(a)$ is shown in Fig.~\ref{fig:beta_vs_f} 
for various values of the fixed radius $R$. 
The curves all have positive slopes, as expected: 
for finer lattices the cluster radius in lattice units $R/a$ has to 
be larger 
so that the radius in physical units $R=a\cdot R/a$ remains fixed.
Thus, for smaller $a$ the clusters must be made larger (in lattice units) via decreasing $f$.
In the following, the interpolation of the $f(a)$ curve will be used 
to set the cut parameter (for a few $N_t=10$ simulation points at 
high temperature, a controlled extrapolation is also necessary).

Having fixed the precise definition of the clusters -- i.e.,\ the 
dependence of the cut parameter on the lattice spacing -- at $T=B=0$, 
we proceed to determine various properties of the 
clusters at nonzero temperatures and nonzero background magnetic fields.

\subsection{Fractality and the mean free path}
\label{sec:Mean_free_path}

We continue the analysis by demonstrating the 
fractal nature of the clusters. To this end, we employ 
the box-counting method to define the fractal dimension $d_\Box$. 
This approach is based on the scaling
\begin{equation}
N(s)\propto s^{-d_{\Box}},
\end{equation}
of the number $N$ of boxes of linear size $s$ necessary to cover a given cluster.
This method was applied and compared to different definitions 
for pure gauge theory in Ref.~\cite{Endrodi:2014yaa}. 

\begin{figure}[t]
    \centering
    \begin{minipage}[]{.4\textwidth}
        \includegraphics[width=\textwidth]{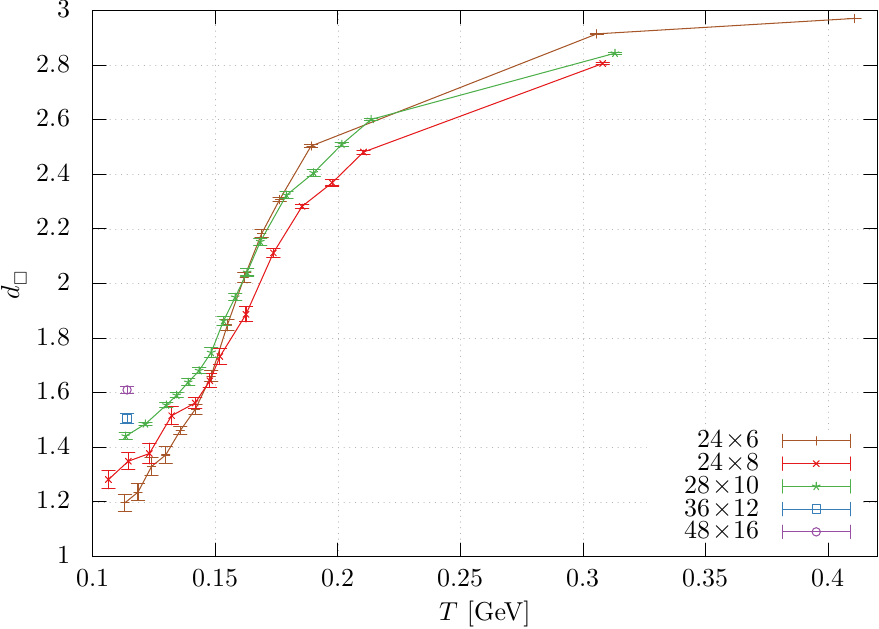}
        \caption
        [Continuum extrapolation of the fractal dimension.]
        {The fractal dimension as a function of the temperature
            for three lattice spacings.
The zero-temperature cluster radius is fixed to $R=0.49\textmd{ fm}$. 
        }
        \label{fig:frac.extr}
    \end{minipage}
\end{figure}

Fig.~\ref{fig:frac.extr} shows the fractal dimension 
of the largest cluster as a function of the temperature for several 
different lattice spacings.
At $T=113 \textmd{ MeV}$, where five lattice spacings are available, the $a\to0$ 
extrapolation gives $d_\Box=1.9(1)$ in the continuum limit. 
For higher temperatures we find that three lattice spacings do not suffice for a 
controlled continuum extrapolation of this observable. 

In center cluster-based models of heavy-ion collisions a relevant 
parameter is the mean free path of partons inside the clusters. It is 
defined as the average distance that the parton can move without scattering 
on the cluster walls. To translate this notion into our setup, 
we consider the following procedure. 
For each site $s$ inside a cluster, we count the number $n_s^i$ of 
sites one can move in the direction $i$
without reaching the boundary of the cluster. 
The (average) mean free path is then given by 
\begin{align}
\lambda_f = \frac{1}{3} \sum\limits_{i=x,y,z}  \lambda_f^{(i)}, \quad\quad
    \lambda_f^{(i)}
    &=
    \frac{1}{S}\, \sum_{s=1}^S \; n_{s}^i \cdot a, 
    \label{eq:mfp}
\end{align}
where $S$ is the total number of sites available for the clusters.

\begin{figure}[t]
    \centering
    \begin{minipage}[]{.4\textwidth}
        \includegraphics[width=\textwidth]{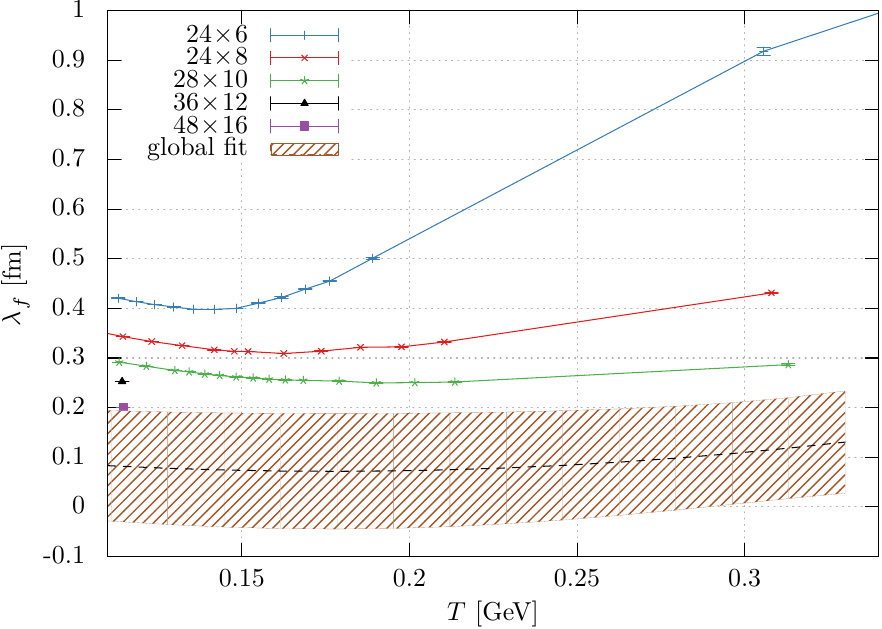}
        \caption
        [Continuum extrapolation of the mean free path.]
        {The mean free path as a function of the temperature, 
            for three lattice spacings and a continuum extrapolation.
        }
        \label{fig:mfp.extr}
    \end{minipage}
\end{figure}

Above we have seen that the clusters are not three-dimensional objects but 
fractals. In the pure gauge theory setting 
it was pointed out already in Ref.~\cite{Endrodi:2014yaa} that as a consequence 
of this fractality the mean free path is not related to the linear cluster size but is 
much smaller than that. 
Using a continuum extrapolation based on three different lattice spacings, 
we show that $\lambda_f$ is consistent with zero for $T\lesssim300\textmd{ MeV}$ in 
the continuum limit, see Fig.~\ref{fig:mfp.extr}.
The systematic error of the continuum extrapolation is estimated by 
comparing fits with different forms for the $T$- and $N_t$-dependence of $\lambda_f$.
On a finite lattice, the fractal pattern is not resolved on distances smaller than the
lattice spacing, thus
the mean free path is bounded from below by $a$. Indeed, Fig.~\ref{fig:mfp.extr} 
reveals how the finite $N_t$ results for $\lambda_f$ approach zero via nonzero values. 
Note that as the temperature is increased further at fixed lattice spacing $a$, 
and the largest cluster becomes 
three-dimensional, $\lambda_f$ will approach half the linear lattice size 
(i.e.,\ it will diverge in the infinite volume limit).

The background magnetic field breaks rotational symmetry and thus might induce an anisotropy 
in the directional mean free paths $\lambda_f^{(i)}$, defined in Eq.~(\ref{eq:mfp}). 
The effect of $B$ on the Polyakov loop (and, thus, on center clusters) is indirect 
and occurs through virtual quark loops. In strong magnetic fields these virtual quarks 
occupy Landau-levels: they are free to move parallel to the magnetic field but are localized 
perpendicular to it. This anisotropy is expected to propagate in the 
gluonic sector and appear in the orientation of center clusters as well, implying 
$\lambda_f^{(z)}>\lambda_f^{(x)}=\lambda_f^{(y)}$. 
Another argument supporting this hierarchy is based on the finding~\cite{Bonati:2014ksa} 
that the magnetic field 
reduces the string tension in the parallel but increases it in the perpendicular direction. 
Indeed, a reduced string tension implies enhanced correlations between distant Polyakov loops 
and, thus, an increased mean free path in a given direction.
Interestingly, in the asymptotically strong magnetic field limit of QCD~\cite{Miransky:2002rp} 
the parallel string tension even vanishes and 
local Polyakov loops are independent of $z$~\cite{Endrodi:2015oba}. Therefore, in this limit 
center clusters become tubes in the $z$-direction but are expected to retain their fractal 
nature in the $x-y$ plane. Nevertheless, our largest available magnetic field 
$eB=3.25\,\textmd{GeV}^2$ is still well below this asymptotic limit.

To determine whether the predicted anisotropy is present in the center structure 
we calculated the directional 
mean free paths at $eB=3.25\,\textmd{GeV}^2$. 
In accordance with the above expectation we observe $\lambda_f^{(z)}$ to exceed 
the perpendicular mean free paths, although only by a few percent. For lower magnetic fields
$0<eB<0.7\,\textmd{GeV}^2$
the effect is found to be smaller than our statistical errors. 
In this range the main effect of the magnetic field turned out to be described by a shift of 
the transition region towards lower temperatures. 
We discuss this effect in more detail in the next section.

\subsection{The QCD phase diagram}
\label{sec:QCD_phase_Diagram}

Due to the crossover nature of the deconfinement transition in full QCD, 
the observables sensitive to the transition exhibit no singular 
behavior but are instead smooth functions of the temperature. 
An implication of this 
is that the transition temperature is not uniquely defined: 
different definitions
may result in different values for $T_c$.

The most straightforward definition involves the inflection point of the average Polyakov loop. 
However, this turns out to be numerically difficult to locate 
due to the slow and gradual rise of $P$ 
with the temperature, cf.\ Ref.~\cite{Bruckmann:2013oba}. 
There have been proposals to circumvent this issue by considering, e.g., ratios of 
Polyakov loop susceptibilities that take well-defined values both well above and 
well below $T_c$, see Ref.~\cite{Lo:2013hla}.

\begin{figure}[t]
         \centering
         \includegraphics[width=.4\textwidth]{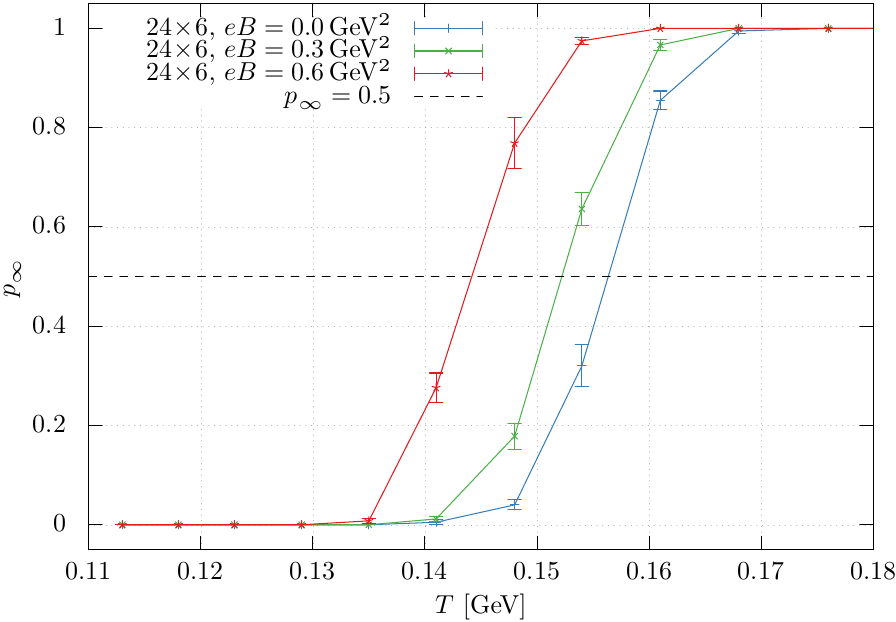}
         \caption[
             Percolation probability, QCD, smeared. 
         ]{The percolation probability as a function of the temperature 
             for three different values of the magnetic field.
             The zero-temperature cluster radius is fixed to $R=0.4\textmd{ fm}$.
         }
         \label{fig:percolation_probability_B_S}
 \end{figure}

Here we propose a new method to define $T_c$ using center 
clusters. 
In terms of the center structure, the most substantial difference between the 
confined/deconfined regimes is the absence/presence of percolating clusters.\footnote{
    Note that this direct realization of the Svetitsky-Yaffe conjecture 
    becomes considerably more involved for $\mathrm{SU}(N)$ theories 
    with $N\ge 4$. Unlike for $\mathrm{SU}(3)$ -- where the Polyakov 
    loop effective action is constructed exclusively via $L(x)$ of 
    Eq.~(\ref{eq:local_ploop}) -- for $\mathrm{SU}(4)$, it involves 
    the trace of gauge links in representations with different dimensions 
    (4 and 6)~\cite{Meisinger:2001cq,Strodthoff:2010dz,Dirnberger:2012gn}. 
    The 4-dimensional representation alone was shown to be insufficient 
    to describe the deconfinement transition via percolation, since 
    the clusters were found to become too thin towards the continuum 
    limit~\cite{Dirnberger:2012gn}. (This is in line with the expectation 
    based on random percolation theory, where the equally populated sectors 
    have probability $p=1/4<p_c$, cf.\ footnote 1.) 
    Here we constrain the discussion to $N=3$, where such complications are absent.
} 
(A cluster is defined to be percolating if it spans across the lattice in 
at least one spatial direction. 
Thus, such clusters become infinitely large in the infinite volume limit.)
The simplest choice reflecting the abrupt change of gluonic 
configurations in this respect is the percolation probability $p_\infty$~\cite{Endrodi:2014yaa,Gattringer:2010ms,Dirnberger:2012gn,Borsanyi:2010cw,Danzer:2010ge}. 
It is defined as the probability of having a percolating cluster
and is thus bounded as $0\le p_\infty \le 1$. 
In Fig.~\ref{fig:percolation_probability_B_S} we plot $p_\infty$ 
as measured on the $24^3\times 6$ lattices, showing the expected rapid increase around $T_c$. 

Taking into account the limiting values of $p_\infty$ at low and at high temperatures, 
respectively,
the most convenient choice for defining $T_c$ is through the implicit 
equation
\begin{align}
    p_\infty (T_c)
    &=
    0.5.
    \label{eq:perc_tc}
\end{align}
This definition will be employed below to map out the phase diagram for 
nonzero magnetic fields. 

\begin{figure}[t]
         \centering
         \includegraphics[width=.4\textwidth]{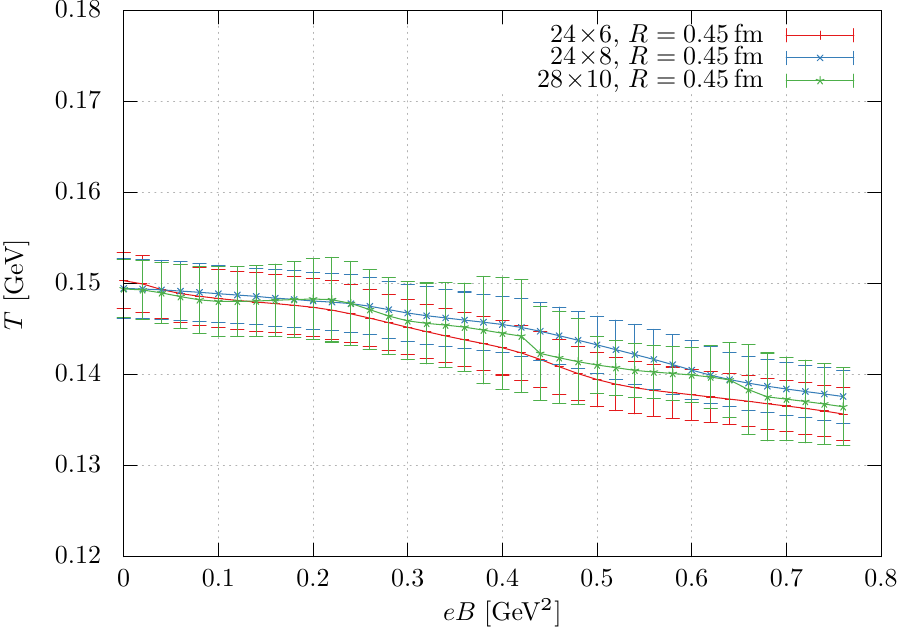}
         \caption[
             Phase diagram, $R=0.45\,$fm.
         ]{The transition temperature, defined according to Eq.~(\ref{eq:perc_tc}), 
	as a function of the magnetic field for three different lattice spacings.
	The zero-temperature cluster radius is fixed to $R=0.45\textmd{ fm}$.
         }
         \label{fig:Tc.R0.45}
 \end{figure}

To demonstrate the effect of magnetic 
fields\footnote{We found that there is no anisotropy in the 
percolation probabilities, even for our strongest magnetic field. Instead, $B$ only induces 
a weak anisotropy over shorter length scales, as revealed by the hierarchy in the directional 
mean free paths discussed in Sec.~\ref{sec:Mean_free_path}.} on $p_\infty$, 
Fig.~\ref{fig:percolation_probability_B_S} also includes the percolation 
probability for a few nonzero values of $B$.
Clearly, the magnetic field 
increases $p_\infty$ for all temperatures and, as a result, reduces the 
transition temperature. This is consistent with previous determinations of 
$T_c(B)$ using chiral quantities~\cite{Bali:2011qj}. 
To quantify this effect, we employed the definition~(\ref{eq:perc_tc}) to 
determine $T_c$ for a range of magnetic fields using three lattice ensembles with 
$N_t=6$, $8$ and $10$. 
Fig.~\ref{fig:Tc.R0.45} shows the so obtained $T_c(B)$, revealing that the results for all 
three lattice spacings fall on top of each other.  The transition temperature is 
found to decrease by about $10\%$ up to $eB=0.75\textmd{ GeV}^2$. 

\begin{figure}[t]
         \centering
         \includegraphics[width=.4\textwidth]{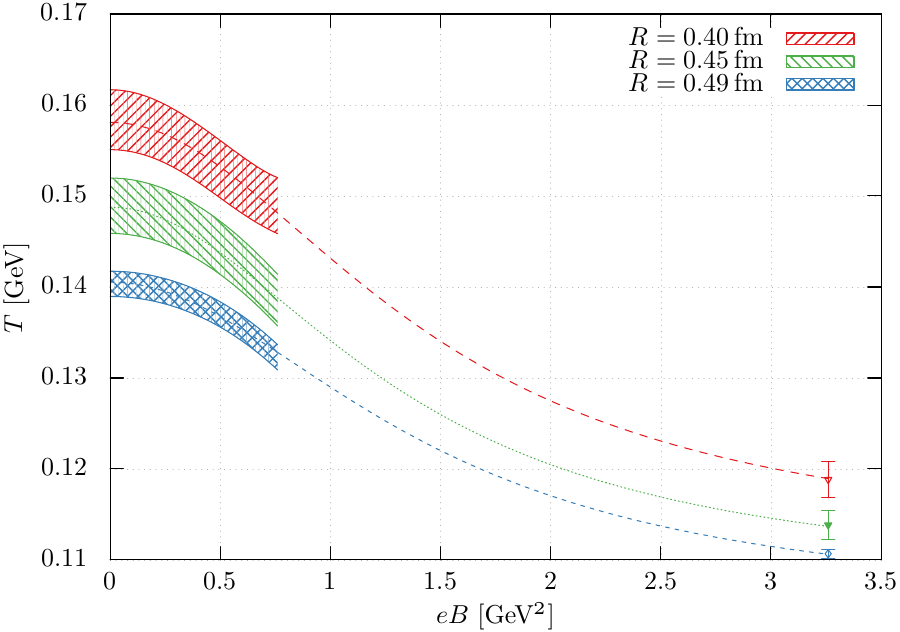}
         \caption[
             Phase diagram, continuum extrapolation, $R=0.40\,$fm, $R=0.45\,$fm, and $R=0.49\,$fm. 
         ]{Continuum extrapolated transition temperatures as a function of the 
             magnetic field for different zero-temperature cluster radii $R$.
             (The last three points on the far right are not continuum extrapolated 
		but were obtained on our $N_t=16$ ensemble.)
         }
         \label{fig:Tc_cluster_continuum}
 \end{figure}

Above, the cut parameter was set by fixing the zero-temperature cluster radius
to $R=0.45\textmd{ fm}$. Also it is of interest how 
the results change if $R$ is varied. We have performed the same analysis for different 
values of $R$. 
Fig.~\ref{fig:Tc_cluster_continuum} shows the continuum extrapolated transition 
temperatures based on our 
three lattice spacings
for three values of the cluster radius, $R=0.40\,$fm, $R=0.45\,$fm, and $R=0.49\,$fm. 
The net effect of decreasing $R$ is to shift the transition temperature 
up. This is to be expected: the smaller the low-temperature clusters are, the 
stronger ordering in the local Polyakov loops (i.e.,\ the higher temperature) is necessary 
for percolation to set in.
Notice that $R$ affects $T_c$ because of the crossover nature of the transition, i.e.,\ 
because the percolation probability depends smoothly on the temperature (even in the 
infinite volume limit). 
The gradual enhancement of $p_\infty(T)$ around $T_c$ 
becomes a real jump in pure gauge theory~\cite{Endrodi:2014yaa}, where 
the transition is of first order. In the latter case, the clusters start to percolate 
suddenly, so that finite changes in the low-temperature cluster radius $R$ are not 
expected to affect $T_c$.
Therefore, the change in $T_c$ due to varying $R$ gives a measure for the width (strength) 
of the deconfinement transition. 

It has recently been shown that the QCD phase diagram exhibits a 
critical endpoint for extremely strong magnetic fields~\cite{Endrodi:2015oba}, 
where the crossover turns into a first order transition (see also Ref.~\cite{Cohen:2013zja}). 
Fig.~\ref{fig:Tc_cluster_continuum} also shows $T_c$ at a very 
large\footnote{This is still well below the estimated critical magnetic field
    $eB_{\mathrm{CEP}}=10(2)\,\mathrm{GeV}^2$~\cite{Endrodi:2015oba}.} 
magnetic field for the three low-temperature cluster radii. 
A further decrease in $T_c$ by about 20\% can be observed, again in agreement with 
previous findings based on other observables~\cite{Endrodi:2015oba}.  
Moreover, the difference between the $T_c$ curves for the different radii 
decreases by about 50\% from $eB=0$ to $eB\approx 3.25\,\mathrm{GeV}^2$. 
According to our reasoning above this shows that the transition becomes 
stronger as the magnetic field grows and the predicted critical point is approached.

\section{Conclusions}
\label{sec:Conclusion}

In this paper we have presented first continuum extrapolated results for 
various observables related to the 
center structure of full dynamical QCD. 
Center clusters were identified using a consistent thinning 
technique involving one parameter (the cut parameter $f$) that is 
fixed by prescribing the cluster radius $R$ at low temperatures.

Using this prescription, 
the fractal dimension of the center clusters was shown to be significantly
smaller than three. We demonstrated that this leads to a vanishing 
mean free path in the cluster structure over the range of temperatures 
$110\textmd{ MeV}<T<300\textmd{ MeV}$. 
We found that the presence of magnetic fields $eB\lesssim 3.25\textmd{ GeV}^2$  does not change 
this result qualitatively -- even at our strongest magnetic field 
the anisotropy in the cluster orientation remains below a few percent.
Thus, for a broad range of temperatures and magnetic fields that are relevant 
for heavy-ion collision phenomenology, the continuum extrapolated mean free path vanishes.
This finding
suggests a limited applicability for models that build on a finite 
mean free path for scattering processes in the QGP.

Furthermore, we proposed a method to define $T_c$ in full QCD 
using the percolation probability and employed this definition to determine 
the phase diagram for nonzero background magnetic fields. 
The results unambiguously show a reduction of $T_c$ with increasing $B$, 
in good agreement with the results obtained using other QCD 
observables~\cite{Bali:2011qj,Endrodi:2015oba}.
In addition, the variation of $T_c$ when changing the zero-temperature cluster 
radius $R$ was argued to measure the width of the crossover transition. 
This quantity was 
found to gradually decrease as $B$ grows and the predicted critical endpoint at 
extremely strong magnetic fields is approached. 
Altogether, our findings demonstrate that the deconfinement transition in full three-color QCD 
can be described as a percolation phenomenon. 
The analysis of further observables and the discussion of finite volume effects 
will be performed in a forthcoming study~\cite{cluster_paper_to_come}.
Finally, we note that generalizations of the percolation picture to
other gauge groups, e.g., $\mathrm{SU}(N)$ with $N>3$,
are non-trivial, and that more extensive research is required
to adress their viability.

\acknowledgments
This work was supported by the DFG (SFB/TRR 55). The authors thank 
Gunnar Bali, Falk Bruckmann, Pavel Buividovich, Christof Gattringer and 
Hans-Peter Schadler for useful discussions.

\bibliography{centerClusterPaper}

\end{document}